\documentclass[reprint, amsmath,amssymb,aps,prb,superscriptaddress]{revtex4-2}

\makeatletter
\newcommand*{\rom}[1]{\expandafter\@slowromancap\romannumeral #1@}
\makeatother

\usepackage{graphicx}
\usepackage{dcolumn}
\usepackage{bm}
\usepackage[usenames,dvipsnames]{color}

\begin{document}

\title{An unconventional platform for two-dimensional Kagome flat bands on semiconductor surfaces}
%%%%%%%%%%%%%%%%%%%%%%%%%%%%%%%%%%%%%%%%%%%%%%%%%%%%%%%%%%%%%%%%%%%%%%%%%%%%%%%%%%%%%%%%%%
\author{Jae Hyuck Lee}\thanks{Both authors equally contributed to this work}\affiliation{Center for Correlated Electron Systems, Institute for Basic Science, Seoul 08826, Republic of Korea}\affiliation{Department of Physics and Astronomy, Seoul National University, Seoul 08826, Republic of Korea} 
\author{GwanWoo Kim}\thanks{Both authors equally contributed to this work}\affiliation{Department of Physics and HMC, Sejong University, Seoul 05006, Republic of Korea}
\author{Inkyung Song}\affiliation{Center for Correlated Electron Systems, Institute for Basic Science, Seoul 08826, Republic of Korea}
\author{Yejin Kim}\affiliation{Department of Physics, Jeonbuk National University, Jeonju 54896, Republic of Korea}
\author{Yeonjae Lee}\affiliation{Center for Correlated Electron Systems, Institute for Basic Science, Seoul 08826, Republic of Korea}\affiliation{Department of Physics and Astronomy, Seoul National University, Seoul 08826, Republic of Korea}
\author{Sung Jong Yoo}\affiliation{Hydrogen·Fuel Cell Research Center, Korea Institute of Science and Technology (KIST), Seoul 02792, Republic of Korea}\affiliation{Division of Energy \& Environment Technology, KIST School, University of Science and Technology (UST), Daejeon 34113, Republic of Korea}\affiliation{KHU–KIST Department of Converging Science and Technology, Kyung Hee University, Seoul 02447, Republic of Korea} 
\author{Deok-Yong Cho}\affiliation{Department of Physics, Jeonbuk National University, Jeonju 54896, Republic of Korea}
\author{Jun-Won Rhim}\email[E-mail: ]{phyruth@gmail.com}\affiliation{Department of Physics, Ajou University, Suwon 16499, Korea}\affiliation{Research Center for Novel Epitaxial Quantum Architectures, Department of Physics, Seoul National University, Seoul, 08826, Korea}
\author{Jongkeun Jung}\email[E-mail: ]{jjk3341@gmail.com}\affiliation{Hydrogen·Fuel Cell Research Center, Korea Institute of Science and Technology (KIST), Seoul 02792, Republic of Korea}
\author{Gunn Kim}\email[E-mail: ]{gunnkim@sejong.ac.kr}\affiliation{Department of Physics and HMC, Sejong University, Seoul 05006, Republic of Korea}
\author{Changyoung Kim}\email[E-mail: ]{changyoung@snu.ac.kr}\affiliation{Center for Correlated Electron Systems, Institute for Basic Science, Seoul 08826, Republic of Korea}\affiliation{Department of Physics and Astronomy, Seoul National University, Seoul 08826, Republic of Korea} 
\email[REVTeX Support: ]{revtex@aps.org}
%%%%%%%%%%%%%%%%%%%%%%%%%%%%%%%%%%%%%%%%%%%%%%%%%%%%%%%%%%%%%%%%%%%%%%%%%%%%%%%%%%%%%%%%%%
\date{\today}
%%%%%%%%%%%%%%%%%%%%%%%%%%%%%%%%%%%%%%%%%%%%%%%%%%%%%%%%%%%%%%%%%%%%%%%%%%%%%%%%%%%%%%%%%%
\begin{abstract}
In condensed matter physics, the Kagome lattice and its inherent flat bands have attracted considerable attention for their potential to host a variety of exotic physical phenomena. Despite extensive efforts to fabricate thin films of Kagome materials aimed at modulating the flat bands through electrostatic gating or strain manipulation, progress has been limited. Here, we report the observation of a novel $d$-orbital hybridized Kagome-derived flat band in Ag/Si(111) $\sqrt{3}\times\sqrt{3}$ as revealed by angle-resolved photoemission spectroscopy. Our findings indicate that silver atoms on a silicon substrate form a Kagome-like structure, where a delicate balance in the hopping parameters of the in-plane $d$-orbitals leads to destructive interference, resulting in a flat band. These results not only introduce a new platform for Kagome physics but also illuminate the potential for integrating metal-semiconductor interfaces into Kagome-related research, thereby opening a new avenue for exploring ideal two-dimensional Kagome systems.
\end{abstract}
\maketitle
%%%%%%%%%%%%%%%%%%%%%%%%%%%%%%%%%%%%%%%%%%%%%%%%%%%%%%%%%%%%%%%%%%%%%%%%%%%%%%%%%%%%%%%%%%
\section{Introduction}
Kagome materials have captivated the scientific community due to their intriguing band topologies and the potential for exotic quantum phenomena. These materials uniquely feature flat bands coexisting with Dirac points and van Hove singularities~\cite{Kagome1}. Despite a wealth of research on Kagome materials~\cite{Kagome1,SFBRev,KagomeRules,KagomeFB1,KagomeFB2,KagomeFB3,KagomeFB4,KagomeFB41,KagomeFB43,KagomeFB44,KagomeFB5,KagomeFB6,KagomeFB7,KagomeFB9,KagomeFB10,KagomeFB8,BK1,BK2}, synthesizing or manipulating true \textit{two-dimensional} Kagome lattices with ideal Kagome flat bands has proven challenging. Moreover, conventional Kagome materials that have been primarily researched for their flat bands have been limited to binary metal magnets T$_m$X$_n$ (T=3d TMs, X: Sn,Ge \& m:n=3:1,3:2,1:1)~\cite{KagomeFB1,KagomeFB2,KagomeFB3,KagomeFB4,KagomeFB41,KagomeFB43,KagomeFB44}, tenary magnets (A(Mn,V)$_6$Sn$_6$, Co$_3$Sn$_2$S$_2$)~\cite{Kagome1,KagomeFB5,KagomeFB6,KagomeFB7,KagomeFB9,KagomeFB10}, van der waals materials (AV$_3$Sb$_5$)~\cite{Kagome1,KagomeFB8} and a series of breathing Kagome chalcogenide halides~\cite{BK1,BK2}. 

Traditional experimental methods such as exfoliation and molecular epitaxy are required to fully explore the two-dimensionality of Kagome materials. However, the synthesis and manipulation of high-quality two-dimensional samples are challenging, limiting the progress in the study of low-dimensional Kagome systems~\cite{MOF}. To overcome these limitations, we introduce a new approach to expand the horizons of two-dimensional Kagome physics. Specifically, we fabricate two-dimensional Kagome systems by depositing monolayer of metal atoms onto semiconductor surfaces. 
Such a method has two significant advantages in that 1)~surface reconstructed structure is inherently \textit{two-dimensional}, and 2)~high degrees of freedom in surface engineering is provided by a plethora of naturally induced surface reconstruction lattices on semiconductor surfaces for various deposition materials. 

\begin{figure}
    \includegraphics[width=8.4cm]{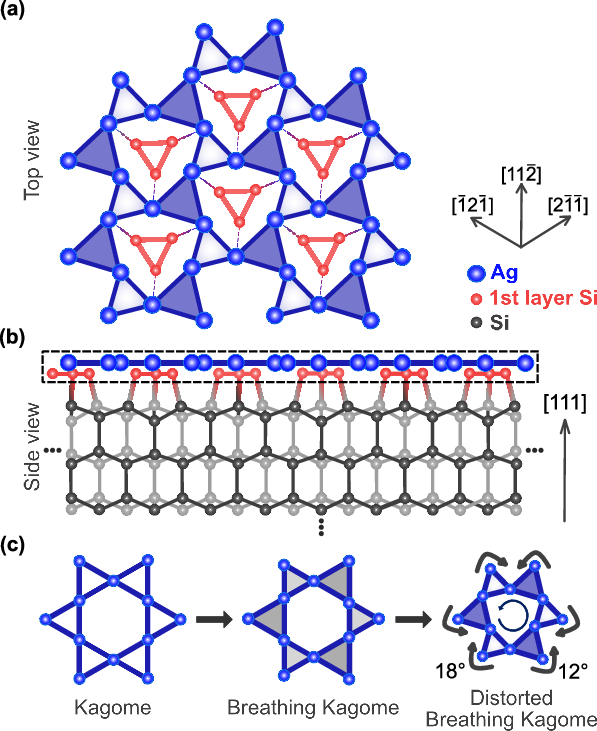}
    \caption{\textbf{Atomic structure and schematic illustration of the Ag/Si system} \textbf{(a)} Top view and \textbf{(b)} side view of the Ag/Si system, respectively. Blue spheres represent Ag atoms, red spheres represent the uppermost Si atoms that form a trimer inside the distorted Ag hexagons, and black and grey spheres represent the bulk Si atoms. \textbf{(c)} Depiction of the distortions leading to the formation of the DBK, starting left from the basic Kagome model to the Kagome lattice with a breathing distortion, and finally to the rotational distortion applied breathing Kagome lattice.}
    \label{fig1}
\end{figure}

As a representative example, we focus on the $\sqrt{3}\times\sqrt{3}$ surface reconstruction of Ag one monolayer on Si(111) substrate, hereafter referred to as the Ag/Si system.
This system has attracted significant attention due to the promising properties of the metal-silicon interface and its potential applications in electronics~\cite{agsi0,agsi1,agsi2}. Although extensively studied, the precise nature of its structure was only recently resolved~\cite{agsi1,agsi2,agsi3,agsi4,agsi5,agsi6,agsi8}  as an "In-Equivalent-Trimer" (IET) configuration~\cite{agsi8}, where the uppermost silver atoms form connected trimers of two different sizes [Fig.~\ref{fig1}(a) and (b)]. Intriguingly, we find that the IET configuration can be reinterpreted as a modified Kagome lattice. As illustrated in Fig.~\ref{fig1}(c), a breathing Kagome lattice is essentially a deformed Kagome lattice with broken inversion symmetry. The IET lattice can be seen as an even further distortion of the breathing Kagome lattice, where the smaller triangle rotates by 18$^\circ$ and the larger one by 12$^\circ$ in the opposite direction. We term this unique structure as a Distorted Breathing Kagome Lattice (DBK).

Using angle-resolved photoemission spectroscopy (ARPES), we confirm the presence of Kagome bands in the Ag/Si system. Subsequent density functional theory (DFT) and tight-binding (TB) calculations confirm the two-dimensional nature of the Ag/Si system and elucidate the origin of the flat band. 
The exotic band structure of the Ag/Si system is explained with a hybridized $d$-orbital Kagome model, which is distinct from a simple $s$-orbital Kagome.
We identify a new type of compact localized state (CLS) and the corresponding destructive interference mechanism from the hybridized $d$-orbital Kagome model. 
We also emphasize that this is the first realization of a flat band Kagome system on Si substrate. As Si hosts numerous surface reconstructed structures depending on the substrate temperature and deposition atoms, our findings shine light on a groundbreaking platform for the study of two-dimensional Kagome systems and open new avenues for future Kagome research on semiconductor surfaces.

%%%%%%%%%%%%%%%%%%%%%%%%%%%%%%%%%%%%%%%%%%%%%%%%%%%%%%%%%%%%%%%%%%%%%%%%%%%%%%%%%%%%%%%%%%%
%%%%%%%%%%%%%%%%%%%%%%%%%%%%%%%%%%%%%%%%%
\begin{figure}
    \includegraphics[width=7.9cm]{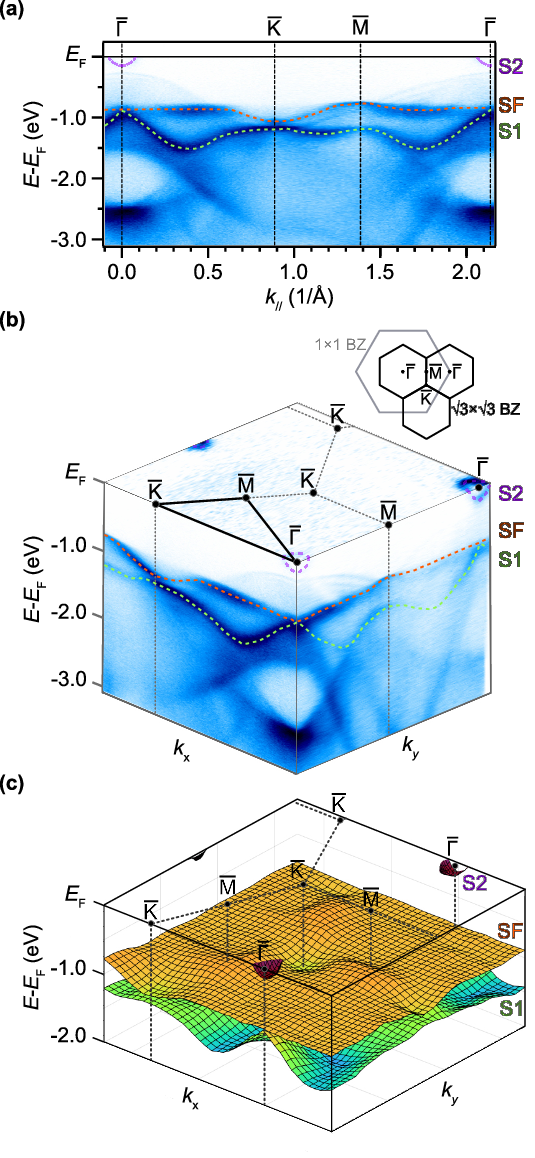}
    \caption{\textbf{ARPES data of the Ag/Si system}. \textbf{(a)} ARPES electronic band structure along the $\overline{\Gamma}-\overline{K}-\overline{M}-\overline{\Gamma}$ cut. \textbf{(b)} 3D ARPES plot of the Ag/Si system. The dotted gray lines represent the BZs with the high-symmetry points denoted. The topmost surface represents the Fermi level 
    ($E_F$). The three surface bands (S1, SF and S2) are highlighted by green, orange, and purple dotted lines, respectively. \textbf{(c)} The three surface bands are fitted and interpolated from (b).}
    \label{fig2}
\end{figure}
%%%%%%%%%%%%%%%%%%%%%%%%%%%%%%%%%%%%%%%%%%%%%%%%%%%%%%%%%%%
\section{Electronic band structure of the Ag/Si system}
We first look into the electronic band structure through the ARPES results displayed in Fig.~\ref{fig2}(a) and (b). Figure~\ref{fig2}(a) is an APRES cut along the $\overline{\Gamma}-\overline{K}-\overline{M}-\overline{\Gamma}$ direction(taken along the thick line in Fig.~\ref{fig2}(b)) while Fig.~\ref{fig2}(b) displays the whole three-dimensional band structure of the Ag/Si system. The hexagonal reciprocal lattice matches the $\sqrt{3}\times\sqrt{3}$ surface structure. For both Fig.~\ref{fig2}(a) and (b), binding energy down to $\approx$-3 eV are plotted to show both Si bulk bands and Ag-Si surface states~\cite{agsi1}. From Fig.~\ref{fig2}(a) and (b) it is shown that there exist three surface bands; a flat band (SF), a dispersive band just below the flat band (S1), and an electron pocket crossing $E_F$ (S2), each represented by orange, green, and purple dotted lines, respectively. These surface bands were assigned as derived from the topmost Ag and Si atoms~\cite{agsi0,agsi1,agsi4}, which will be verified by DFT calculations in the next section. Further TB calculations in the next section indicate that all three surface bands have their origin in the Ag Kagome lattice. 

Bands SF and S1 are both located near a binding energy of -1 eV. The SF band remains flat throughout most of the Brillouin Zone (BZ) but shows dispersion ($\lesssim$ 0.15 eV) near the $\overline{K}$ point, as seen in the orange dotted line in Fig.~\ref{fig2}(a) and (b). The broadening of the flat band near the $\overline{\Gamma}$ point is possibly due to hybridization with the weak hole-like bands that lie between S2 and SF at the $\overline{\Gamma}$ point. Near the BZ boundary, on the other hand, the flat band is located above the hole-like bands in the bulk-gap region, making it sharper. These weak but distinct hole-like bands are attributed to Si-related surface resonance bands, hole subbands in a Si quantum well, or both~\cite{agsi4,QWSI}.
 
In addition to the flat band, the ARPES result confirms another signature of a Kagome band: a quadratic band touching (QBT) at the $\overline{\Gamma}$ point, which is a key characteristic of a Kagome band structure~\cite{SFBRev, QBT, FBloopTop}. 
A flat band having this type of QBT is called a singular flat band~\cite{SFBRev}, which is expected to show many exotic quantum geometric phenomena characterized by the quantum distance, such as the anomalous Landau levels~\cite{QUANTUMDISTANCE}, and the unconventional bulk-boundary correspondence~\cite{bulkInterQuantDis}.
We note that although there should be a spin-orbit coupling (SOC)-induced gap between SF and S1 at the $\overline{\Gamma}$ point, it is hard to identify within the resolution of the ARPES data due to band broadening near the $\overline{\Gamma}$ point. Moreover, the gap size is small($\approx$16 meV) due to the weak SOC strength of Ag [see Supplemental Materials Fig.~S2]. However, by replacing Ag with heavy elements with strong SOC, we expect to realize a topological nearly flat band~\cite{QUANTUMDISTANCE}. 

Finally, S2 forms an isotropic electron pocket with a minimum binding energy (E$_B$) of approximately -0.15 eV at the $\overline{\Gamma}$ point. This metallic S2 band forms a circular Fermi surface, indicating two-dimensional free-electron-like surface states, and has been relatively well studied compared to other surface bands~\cite{agsi0,agsi9,hirahara}. 

For clarity, only the three surface bands SF, S1 and S2 were fitted and extrapolated to surfaces for the whole BZ and are plotted in 3D in Fig.~\ref{fig2}(c). The S2 electron pocket is revealed as the closed purple surface centered around the $\overline{\Gamma}$ point. As hinted in Fig.~\ref{fig2}(a) and (b), the flat band of SF is clearly visualized across most of the BZ, with a small degree of dispersion on the zone boundary. The flatness and zone boundary dispersion of SF are directly realized by the surface color gradient of SF in Fig.~\ref{fig2}(c), where the largest change is at the surface dip near the $\overline{K}$ point (scaling to $\lesssim$ 0.15 eV). The dispersion near the $\overline{K}$ point is driven by next-nearest-neighbor hopping, which is typically found in realistic Kagome models~\cite{QUANTUMDISTANCE}. Compared to SF, S1 is highly dispersive and crosses with SF only at the $\overline{\Gamma}$ point via QBT, which we emphasize again is of a Kagome nature.

\section{Origin of the flat band}
\subsection{DFT calculations} 
%%%%%%%%%%%%%%%%%%%%%%%%%%%%%%%%%%%%%%%%%%%%%%%%%%%%%%%%%%%%%
\begin{figure*}
    \includegraphics[width=17.5cm]{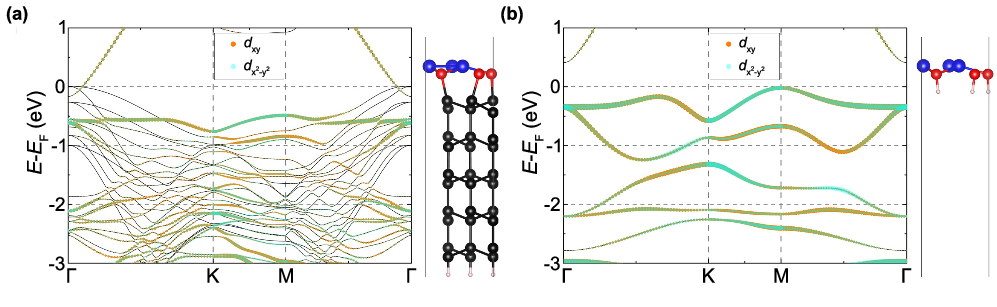}
    \caption{\textbf{Layer-dependent DFT calculations}
    \textbf{(a)} Fat-band calculations for hydrogen-terminated slab system with multi-layer Si atoms. 
    \textbf{(b)} The thickness of the slab system has been dramatically reduced leaving only the topmost Si and Ag layer. 
    On the left panels of (a) and (b), the fat-band calculations for the 4$d_{xy}$ and 4$d_{x^2-y^2}$ orbitals of Ag atoms are displayed. The orbital contributions of the 4$d_{xy}$ and 4$d_{x^2-y^2}$ orbitals of Ag atoms are shown by the size of the orange and cyan circles, respectively. Side-view of the atomic structure for each of the slab system are depicted on the right panel. Ag, topmost Si, bulk Si, and H atoms are represented by blue, red, black, and white spheres respectively. 
    } 
    \label{fig3}
\end{figure*}
%%%%%%%%%%%%%%%%%%%%%%%%%%%%%%%%%%%%%%%%%%%%%%%%%%%%%%%%%%%%%
To investigate the origin of the observed flat band in the Ag/Si system, we employed DFT calculations. Our computational model, which was carefully designed to match the experimental conditions, helped us to map the flat bands onto the atomic orbitals within the structure.
The DFT results, displayed in Fig.~\ref{fig3}(a), demonstrate excellent agreement with our ARPES data. The accuracy of our DFT calculations is shown by the fact that they were able to reproduce the three surface bands (SF, S1, and S2). Moreover, the hole subbands between the S2 and SF band and bulk bands from the Si substrate are all present. However, it is difficult to understand the complex nature of these bands by simply looking at them. We need to use a more sophisticated approach.

Our analysis suggests the flat bands likely originate from the surface of the material. Projected density of states (DOS) calculations reveal that the dominant contribution to the surface bands emanates from the in-plane Ag atomic orbitals, specifically the 4$d_{xy}$ and 4$d_{x^2-y^2}$ orbitals, shown in orange and cyan in Fig.~\ref{fig3}(a) and \ref{fig3}(b), respectively. Additional contributions are observed from Si $p$-orbitals.

To gain deeper insights into the complex interactions of these hybrid orbitals, we needed a simplified model. We, therefore, performed additional calculations with fewer layers of Si atoms to verify the primary contributions of the surface bands from the uppermost atoms, particularly the Ag atoms in the DBK lattice. This allowed us to focus solely on the contributions from Ag and Si atoms at the surface.
Interestingly, when we kept the Ag layer and removed most of the silicon layers, the flat bands remained. In contrast, the Si bulk and other bands disappeared, revealing a gapped Dirac band-like feature characteristic of a breathing Kagome lattice between -0.4 eV and -2.2 eV, as shown in Fig.~\ref{fig3}(b). Notably, a novel flat band was observed around -2.2 eV in the single-layer DFT calculation, intersecting the Dirac band at a QBT point. This additional flat band at -2.2 eV raises an intriguing question about the limitation of the $s$-orbital Kagome lattice model in describing the $d$-orbital Kagome system. Further investigation on this is desired, focusing on the mechanism and its potential application.

Tracing the behavior of the Dirac-like band was difficult in our original model because it interacted with the Si bulk band. However, by decreasing more Si layers [see Supplemental Materials Fig.~S3], we could see how the band evolved. Even though the band changed slightly, the 4$d_{xy}$ and 4$d_{x^2-y^2}$ orbitals remained within the Kagome-like band, confirming that this band persisted and continued to be influenced by these Ag orbitals. Additionally, as we removed more Si layers, the orbital contributions to the flat bands shifted, further highlighting the importance of the in-plane Ag orbitals.

%%%%%%%%%%%%%%%%%%%%%%%%%%%%%%%%%%%%%%%%%%%%%%%%%%%%%%%%%%%%%
\subsection{Tight-Binding analysis and origin of surface bands}
\begin{figure*}
    \includegraphics[width=15.5cm]{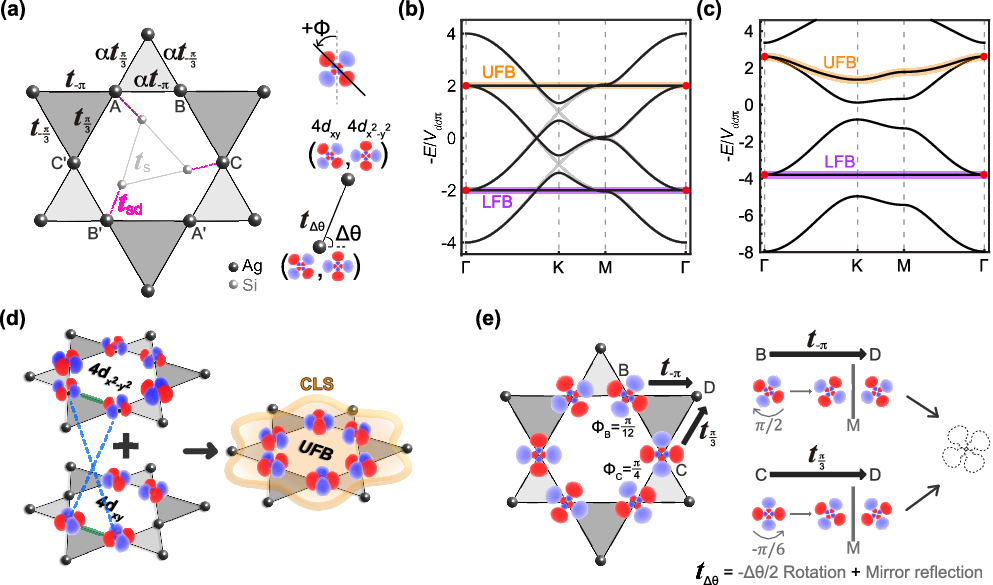}
    \caption{\textbf{Origin of the flat band}. \textbf{(a)} Schematic of the lattice structure of the Ag/Si system for TB calculation where the breathing Kagome lattice of Ag atoms couples with a Si-triangle inside its hexagon. The hopping matrices $t_{\Delta\theta}$ between the Ag atoms are displayed together with the asymmetry parameter $\alpha$. Grey lines represent hopping between the topmost Si atoms ($t_s$) and the dotted magenta lines stand for the hopping between Ag and Si atoms ($t_{sd}$). The $t_{\Delta\theta}$ consists of intra- and inter-orbital hopping between 4$d_{xy}$ and 4$d_{x^2-y^2}$ (bottom-right schematic). $\phi$ represents the real rotation angle of the $d_{xy}$ orbital (upper-right schematic). \textbf{(b)} Band structure of the multi-orbital Kagome lattice with $V_{dd\pi}=1$, $E_0=0$, $\alpha=1$ (light grey line) and $\alpha=0.8$ (dark black line).  \textbf{(c)} The recreated band structure of the Ag/Si system obtained from band parameters $V_{dd\sigma}/V_{dd\pi}=V_{dd\delta}/V_{dd\pi}=-1$, $V_{sd}/V_{dd\pi}=2$, $t_s/V_{dd\pi}=-1$, $\epsilon/V_{dd\pi} = 5$, and $\alpha=2.8$. The UFB and LFB are each emphasized by orange and purple lines respectively, with the QBT marked by a red dot. \textbf{(d)} Schematic of the orbital couplings leading to the formation of CLS in the multi-orbital Kagome lattice ($\alpha$ = 0.8). Intra-orbital hopping and inter-orbital hopping between 4$d_{xy}$ and 4$d_{x^2-y^2}$ are each represented by solid green and broken blue lines respectively. The orbital configuration of the UFB is exhibited and its CLS is represented by the orange shaded hexagon. \textbf{(e)} Mechanism of destructive interference in the UFB.}
    \label{fig4}
    \end{figure*}

To corroborate our layer-dependent DFT results and to elucidate the origin of the flat band with the destructive interference mechanism, we perform a TB analysis on a generalized Kagome structure of the Ag/Si system, as illustrated in Fig.~\ref{fig4}(a). We select only Ag 4$d_{xy}$, 4$d_{x^2-y^2}$, and Si 3$s$ orbitals as the basis of the TB Hamiltonian based on the DFT results. We introduce a vectorial representation  $\textbf{d}=(\alpha,\beta)$ for the linear combination of $d$-orbitals, $|\psi\rangle=\alpha|d_{xy}\rangle+\beta|d_{x^2-y^2}\rangle$. The combined state $\psi$ can be denoted by an angle $\theta$ of $\textbf{d}$ with respect to the $x$-axis [lower right schematic in Fig.~\ref{fig4}(a)]. Note that the real rotation angle $\phi$ of the $d_{xy}$ orbital is half of $-\Delta\theta$ [upper right schematic in Fig.~\ref{fig4}(a)]. For example, the $d_{x^2-y^2}$ orbital, corresponding to $\theta=\pi/2$, is rotated $\phi=-\pi/4$ from the $d_{xy}$ orbital. Then, the hopping of $d$-orbitals ($\textbf{d}$) to the neighboring sites at an angle $\Delta\theta$ is given by $t_{\Delta\theta}=V_{dd\pi}M_{x}R_{\Delta\theta}$, where $M_{x}=-\sigma_{z}$ corresponds to the mirror reflection with respect to the $y$-axis ($x\rightarrow-x$) and $R_{\Delta\theta}$ is a rotation matrix. In our model, only three angles, $\Delta\theta=\pi/3$, $-\pi/3$, and $-\pi$, are used [left schematic in Fig.~\ref{fig4}(a)].

First, we ignore the Si atoms in Fig.~\ref{fig4}(a) ($t_{sd}$=0 and $t_{s}$=0) to only account for the DBK lattice of the Ag atoms. 
The band structures of ideal ($\alpha$=1) and breathing ($\alpha$=0.8) Kagome models are plotted in Fig.~\ref{fig4}(b), where $\alpha$ is an asymmetry parameter between two triangles with different sizes. The breathing Kagome lattice is characterized by the gap opening at the K point due to the breaking of inversion symmetry~\cite{BK1}. However, for both cases, the upper flat band (UFB) and lower flat band (LFB) persist, which are connected to the nearly flat bands found in DFT and experimental results. Also, the QBTs with the quadratic Dirac band at the $\Gamma$ point, protected by the $C_3$ symmetry of the DBK lattice, are present for both cases.

Next, we turn on the coupling of the Ag $d$-orbitals to the Si $s$-orbital ($t_{sd}$$\neq$0) and the hopping between Si ($t_s$$\neq$0), to investigate the deformation of flat bands into nearly flat bands of the realistic system. 
Remarkably, the overall characteristics of the electronic band still remain: both UFB and LFB exist along with the two Dirac bands between the energy range of the two flat bands[see Fig.~\ref{fig4}(c)]. Note that the QBTs are also robustly conserved for both UFB and LFB at the $\Gamma$ point due to the protected $C_3$ symmetry by the triangular Si atoms. This result is in parallel to the layer-dependent DFT band dispersion which took into account only the topmost Ag and Si atoms in Fig.~\ref{fig3}(b). While UFB remains flat near the $\Gamma$ point, dispersion becomes the largest at the K point maintaining a gap between the Dirac band below. 

The uppermost quadratic Kagome band in Fig.~\ref{fig4}(b), on the other hand, undergoes a drastic change after the inclusion of Si orbitals forming an isotropic electron pocket, mimicking the S2 band, as is manifest in Fig.~\ref{fig4}(c). Such modification of the band after the insertion of Si is in line with the previous interpretation that S2 is a surface state having bonding character from both the Ag-Si orbitals~\cite{agsi0,agsi9,agsi1,hirahara}. The electron pocket is also evident in the topmost layer DFT calculation [see Fig.~\ref{fig3}(b)].

Overall, we have shown that the most essential features of the ARPES band structure, such as the nearly flat bands, the free-electron-like-band, and the QBT at $\Gamma$ point, are captured by the electronic structure of the topmost Ag and Si layer from DFT and TB analysis [compare Fig.~\ref{fig2}(a), Fig.~\ref{fig3}(b), and Fig.~\ref{fig4}(c)]. The similarities between the ARPES data, DFT, and TB calculations strongly suggest that the nearly flat bands originate from the Kagome structure of the Ag/Si system. Surprisingly, studies so far regarding hybridized $d$-orbital Kagome systems have not explicitly utilized the double band structure given in Fig.~\ref{fig4}(b) as applied here to the Ag/Si system.

%%%%%%%%%%%%%%%%%%%%%%%%%%%%%%%%%%%%%%%%%%%%%%%%%%%%%%%%%%%%%%%%%%%%%%%%%%%%%%%%%%%%%%%%%%%%%%%%%%%%%
%%%%%%%%%%%%%%%%%%%%%%%%%%%%%%%%%%%%%%%%%%%%%%%%%%%%%%%%%%%%%%%%%%%%%%%%%%%%%%%%%%%%%%%%%%%%%%%%%%%%%
\subsection{Mechanism of the destructive interference} 
The origin of a flat band is understood from the existence of an exotic localized eigenstate CLS, which has finite amplitudes only inside a finite region due to destructive interference~\cite{SFBRev,QBT}. One of the most well-known CLSs is the hexagon-shaped one of the $s$-orbital Kagome lattice model~\cite{QBT2}. While the Ag/Si system has also turned out to be a Kagome structure, we show that the mechanism behind the formation of a flat band is completely different from that of the popular $s$-orbital Kagome model. We demonstrate that the CLS of the Ag/Si system is stabilized by the inter-orbital destructive interference between two $d$-orbitals, as shown in Fig.~\ref{fig4}(d). 

Let us see how the mixed $d$-orbitals at B and C sites can be canceled out at the D site via destructive interference after the hopping processes. At B and C sites, $\theta$ for the orbital compositions of the CLS are given by $\theta_{B}=-\pi/6$ and $\theta_{c}=-\pi/2$ (or $\phi_{B}=\pi/12$ and $\phi_{C}=\pi/4$), respectively. After the hopping processes, $d$-orbitals at B(C) site hop to the neighboring D site via $t_{-\pi}$($t_{\pi/3}$), resulting in $\theta_{B}\rightarrow\theta_{B}+\Delta\theta=-7\pi/6$ or $\phi_{B}\rightarrow\phi_{B}+\Delta\phi=7\pi/12$ ($\theta_{C}\rightarrow\theta_{C}+\Delta\theta=-\pi/6$ or $\phi_{C}\rightarrow\phi_{C}+\Delta\Phi=\pi/12$). Namely, $\phi_{B}$ and $\phi_{C}$ becomes out of phase ($\pi/2$) at the D site after each hopping, implying that their linear combination vanishes and a destructive interference is achieved. 

Such a mechanism is consistent for CLSs in all multiple $d$-orbital-based (breathing) Kagome systems regardless of the minute distortion as seen in the Ag/Si system. Thus we have deduced the CLS corresponding to the flat band SF and have elucidated the underlying quantum destructive interference typically present in Kagome systems.

Another interesting common feature between Fig.~\ref{fig3}(b) and Fig.~\ref{fig4}(c) is that the LFB remains robustly flat compared to the UFB. This is because the CLS of the LFB remains the same even after the coupling with Si atoms is turned on. As shown in detail in Supplemental Materials[see Fig.~S7 and equation (20)], the amplitudes corresponding to the two $d$-orbitals of the LFB at the Ag atoms connected to the Si orbital have opposite signs with ratio inversely proportional to the $sd$-coupling strength. Consequently, the $d$-orbitals cancel each other at the Si sites after hopping. 
We note that the LFB is not apparent in the ARPES data, possibly due to overlap with the complex Si bulk band, and the fact that features are generally not clear at high binding energies.

\section{Summary and Concluding remarks}
\label{Summary}
In summary, we have reinterpreted the Ag/Si system as a new type of Kagome lattice with hybridized $d$-orbitals. Using ARPES, we have measured a flat band with a QBT  at the $\overline{\Gamma}$ point in the surface-reconstructed Ag/Si system. DFT calculations confirm that the dominant orbital contribution of the flat band comes from the in-plane 4$d$-orbitals. Comparisons with TB calculations and DFT on the topmost atomic layers reveal that the flat bands are multi-orbital Kagome bands with quantum destructive interference resulting from a delicate balance between the inter-orbital hopping parameters and the unique bonding geometry from in-plane orbitals. This destructive interference is crucial for non-trivial CLS. 
 
Furthermore, we propose that the UFB energy can be controlled via hole-doping. For flat bands to give rise to unconventional correlation phenomena, they must be close to the $E_F$. Indeed, when the Ag/Si system was grown on a p-doped Si we were able to confirm that the $E_F$ is lowered by $\approx$0.1 eV compared to the one grown on a n-doped Si[see Supplemental Materials Fig.~S9]. We thus propose that further investigations of doping methods could bring the UFB closer to the $E_F$, opening the doors to exotic transport properties and applications. Our result paves the way for a whole new approach to finding Kagome-like bands and also opens up a wide platform where ideal two-dimensional Kagome lattices can be realized.

%%%%%%%%%%%%%%%%%%%%%%%%%%%%%%%%%%%%%%%%%%%%%%%%%%%%%%%%%%%%%%%%%%%%%%%%%%%%%%%%%%%%%%%%%%%%%%%%%%%%%
\section{Methods}
\textit{Sample preparation and ARPES measurements}: Sample fabrication and ARPES measurements were conducted at a home-built in-situ ARPES cluster system in Seoul National University (SNU). The details on the fabrication process is given in Supplemental Materials. A He-\rom{1}$\alpha$ photon source ($h\nu$ = 21.2 eV) emitted from a discharge lamp (Fermion Instrument) was used and the spectra was obtained at 10 K using a Scienta DA30 analyzer. Sample preparation, transfer and measurement were all performed in a UHV environment better than 5$\times$$10^{-11}$ Torr.

\textit{DFT calculations}:
This study used the Vienna Ab initio Simulation Package (VASP) program to perform DFT calculations~\cite{DFT-KS1,DFT-KS2}. The projector-augmented wave (PAW) method~\cite{DFT-PAW} was used to simulate the atomic potential. The metallic nature of the system was confirmed, and the Methfessel-Paxton approximation was used to sample the BZ~\cite{DFT-MP}. A cutoff energy of 600 eV and an 18 $\times$ 18 $\times$ 1 k-grid were used for the electronic structure calculations. The Grimme-D3 method~\cite{DFT-VdW_D3} was used to account for interatomic van der Waals interactions, and the generalized gradient approximation (GGA) functional~\cite{DFT-GGA_PBE} was used for the exchange-correlation functional. Self-consistent calculations were performed with an electronic threshold of 10$^{-6}$ eV and a convergence criterion of 10$^{-3}$ eV/\AA ~for the Hellmann-Feynman forces~\cite{DFT-HF_Force}.

%%%%%%%%%%%%%%%%%%%%%%%%%%%%%%%%%%%%%%%%%%%%%%%%%%%%%%%%%%%%%%%%%%%%%%%%%%%%%%%%%%%%%%%%%%%%%%%%%%%%%
\begin{acknowledgments}
The work by J.H.L., Y.L., and C.K. were supported by the Institute for Basic Science (IBS) (Grant No. IBS-R009-G2) and the National Research Foundation of Korea (NRF) grant funded by the Korean government (MSIT) (Grant No. NRF-2022R1A3B1077234). I.S. was also supported by IBS (Grant No. IBS-R009-G2). GW.K. and G.K. were funded by the NRF (Grant No. NRF-2020R1A6A1A03043435). J.J. and S.J.Y. were supported by the NRF (Grant No. NRF-2021M3H4A1A02042948). J.-W.R. was funded by the NRF (Grant Nos. NRF-2021R1A2C1010572, NRF-2022M3H3A106307411, and NRF-2021R1A5A1032996).
\end{acknowledgments}
%%%%%%%%%%%%%%%%%%%%%%%%%%%%%%%%%%%%%%%%%%%%%%%%%%%%%%%%%%%%%%%%%%%%%%%%%%%%%%%%%%%%%%%%%%%%%%%%%%%%%

\end{document}